\documentclass{svproc}
\usepackage{url}

\usepackage{graphicx}
\usepackage{caption}
\usepackage{comment}
\usepackage{dblfloatfix} 
\usepackage[colorlinks=true,
            linkcolor=red,
            urlcolor=blue,
            citecolor=blue]{hyperref}
\usepackage{subcaption}
\usepackage{comment}
\begin{document}
\mainmatter              
\title{Toward a Knowledge Discovery Framework for Data Science Job Market in the United States}
\titlerunning{Knowledge Discovery Framework}  
%
\author{Mojtaba Heidarysafa\inst{1} \and Kamran Kowsari\inst{1,2} \and
Masoud Bashiri\inst{1}  \and \\Donald E. Brown\inst{1,3}}
\authorrunning{Heidarysafa et al.} 
%
\tocauthor{Ivar Ekeland, Roger Temam, Jeffrey Dean, David Grove,
Craig Chambers, Kim B. Bruce, and Elisa Bertino}
\institute{Department of Systems and Information Engineering, University of Virginia
\and
Office of Health Informatics and Analytics, University of California, Los Angeles
\and
School of Data Science, University of Virginia
}

\maketitle              

\begin{abstract}
The growth of the data science field requires better tools to understand such a fast-paced growing domain. Moreover, individuals from different backgrounds became interested in following a career as data scientists. Therefore, providing a quantitative guide for individuals and organizations to understand the skills required in the job market would be crucial. This paper introduces a framework to analyze the job market for data science-related jobs within the US while providing an interface to access insights in this market. The proposed framework includes three sub-modules allowing continuous data collection, information extraction, and a web-based dashboard visualization to investigate the spatial and temporal distribution of data science-related jobs and skills. The result of this work shows important skills for the main branches of data science jobs and attempts to provide a skill-based definition of these data science branches. The current version of this application is deployed on the web and allows individuals and institutes to investigate skills required for data science positions through the industry lens~\footnote[1]{Version 1.1 of this work is available at \url{https://dsi-usa2.herokuapp.com/}}~\footnote[2]{The code is shared at \url{https://github.com/mojtaba-Hsafa/data-science-jobs-app}}.
\end{abstract}
\section{Introduction}
In 2012, Harvard business review published an article ``Data Scientist: The Sexiest Job of the 21st Century'' and in it, authors predicted that certain sectors will face a shortage of data scientists in the near future\cite{davenport2012data}. Soon after, universities were scrambling to design programs for data science to appropriately fill the gap. However, the term ``Data Science'' is confusing and without a proper understanding, it will fade away into a new buzz word. Although data science is claimed to be driven from statistics as it uses statistical methods~(exploratory analysis, machine learning, reproducibility, etc), the two fields are not the same and in a sense ``Data Science'' encapsulates a broader scope\cite{carmichael2018data}. Data science is more intertwined with other important fields such as big data, and artificial intelligence and often deals with heterogeneous and unstructured data such as text, image and video~\cite{dhar2013data},\cite{provost2013data}.

Another interesting aspect of this new field is related to the growth rate of data science-related jobs. According to linkedin economic graphic, machine learning engineering, and data science were the top emerging jobs between~$2012-2017$ with a growth rate of $9.8X$ and $6.5X$ respectively~\cite{linkedin2017}. According to Glassdoor it was the number one best job in the United States in 2019 and the third one in~$2020$. The expectation is that the need will still increase but the exact growth percentage is not known. These reports indicate that data science is particularly a fast pace growing field with new technologies being introduced to it every day. Therefore, it is important to observe the job market to stay agile in such a fast pace field. 

Other motivation to introduce a framework such as the one presented here includes figuring out different branches of data science as a broad term in job search engines and different skills associated with each of these categories.  Mostly,individuals can not have fluency in all of these skills and thus prioritizing the skills needed for an individual based on the path they want to pursue as a data scientist would be crucial. 

This paper presents a framework to understand data science based on its job market available positions. The proposed framework allows to both look into temporal and spatial changes of data science job market in the United States and the scope of the field and related skills needed for job placements. Moreover, using this approach, the hope is to get a skill base definition of  data science through the industrial requirement lens. The rest of this paper is organized as follow:
Section~\ref{sec:related} presents the related works, Section~\ref{sec:method} discusses our method and Section~\ref{sec:results} presents the result. Finally, Section~\ref{sec:Conclusion} discusses future works. 
\section{Related Works}\label{sec:related}
In this paper, we combined multiple data science tools to build a system capable of understand spatial and temporal distributions of data science skills in the US. Before delving into our  approach, we investigated solutions to related problems from different perspectives. In the first section, we describe the work on market analysis and more specifically data science market analysis. Next, text mining methods for job advertisements will be discussed.

\subsection{Job Market Analysis }
During the past decades, the internet has gradually become the first place to look for jobs and as a result, more and more job advertisements appear on the web. These collections of job ads themselves have become valuable resources in order to investigate what requirements different jobs might have and how industries generally perceive those positions.
As an example, Daneva et al. used online job market advertisements in the Netherlands to understand what industry means by ``requirements engineers''~\cite{daneva2017job}. Although they had a small corpus and did not utilize text mining methods, they were able to present skills, competencies, and preferred background for these positions. Other researchers used job advertisements to investigate trends in the qualifications and responsibilities of Electronic Resources Librarians in a period between 2000–2012~\cite{hartnett2014nasig}. At the same time, the importance of this information-rich source for the industry itself has not been ignored. Multiple companies and start-ups shaped around gathering and creating insights from the job market data. The work described in this paper aims to provide a platform for both finding a skill based definition and understanding the trends in the data science job market.
\subsection{Data Science Market Analysis}
Data science has a unique position among the IT jobs because of particular characteristics of the field. First, data scientists are required to have a very wide range of skills, and usually come from different backgrounds and are recruited by different sectors. On top of that, the field is a fast-moving field with new tools and technologies introduced to it each day. Therefore, it is reasonable for researchers to investigate this domain further and the result could help in understanding the field.

In an article published in 2017 by IBM researcher \textit{Steven Miller}, the authors addressed how disruptive data science is in the labor market. They analyzed job market advertisements in all categories of Data Science jobs in~$2015$ for both skillsets and recruiting sectors~\cite{miller2017quant}. They also made an analysis of the growth rate of skills and projected the number of jobs for the coming years. As previously mentioned, this information can be used to guide educational institutes to build a better curriculum for their data science degrees. \textit{Manieri et al.} took this approach in identifying the needed skills based on a corpus of~$2500$ job ads on data science in~$2015$. They used principal component analysis on their data for dimension reduction and their result identified programming skills, big data skills, database knowledge ,and machine learning as the main components that emerged from data science job markets~\cite{manieri2015teaching}. Moreover, researchers combined courses offered and job advertisements for data science in an effort to develop semi-automatic service to analyze and detect the gaps in demand and supply of skills~\cite{belloum2019bridging}. Although these researches tackled different problems in the data science domain, the work presented here is different from them due to its emphasis on the temporal and spatial aspect of this market and skills and providing a visual quantitative guide for individuals .

\subsection{Text Mining Methods for Job Ads}
Since job posting provides a rich body of information for market analysis, researchers have used it for understanding job markets around the world. A fruitful analysis of these postings however requires applying some sort of text mining methods to these collections. In particular, matching jobs and skills were of interest to these researches. Such skill requirements matching has been researched in different domains such as big data~\cite{gardiner2018skill}, information technology~\cite{wowczko2015skills}, and librarians~\cite{yang2016current} to name a few. The text mining methods in these research vary widely from simple frequency counts of terms to using external resources in conjunction with the job postings. For example,~\cite{wowczko2015skills} used term frequency of terms and after cleaning the corpus from spare words presented the top 30 words in both job titles as well as job descriptions. In~\cite{darabi2018detecting}, the author used a demand skill index which is derived by normalization of words' term frequencies. In another research, they used term frequency - inverse document frequency to rank the top terms~\cite{maer2019skill}. A problem with this approach is that terms are not necessarily equivalent to the skills and there would be a need for post-processing of the result to remove other noises.
\begin{figure*}[!b]
\centering
\includegraphics[width=1\textwidth]{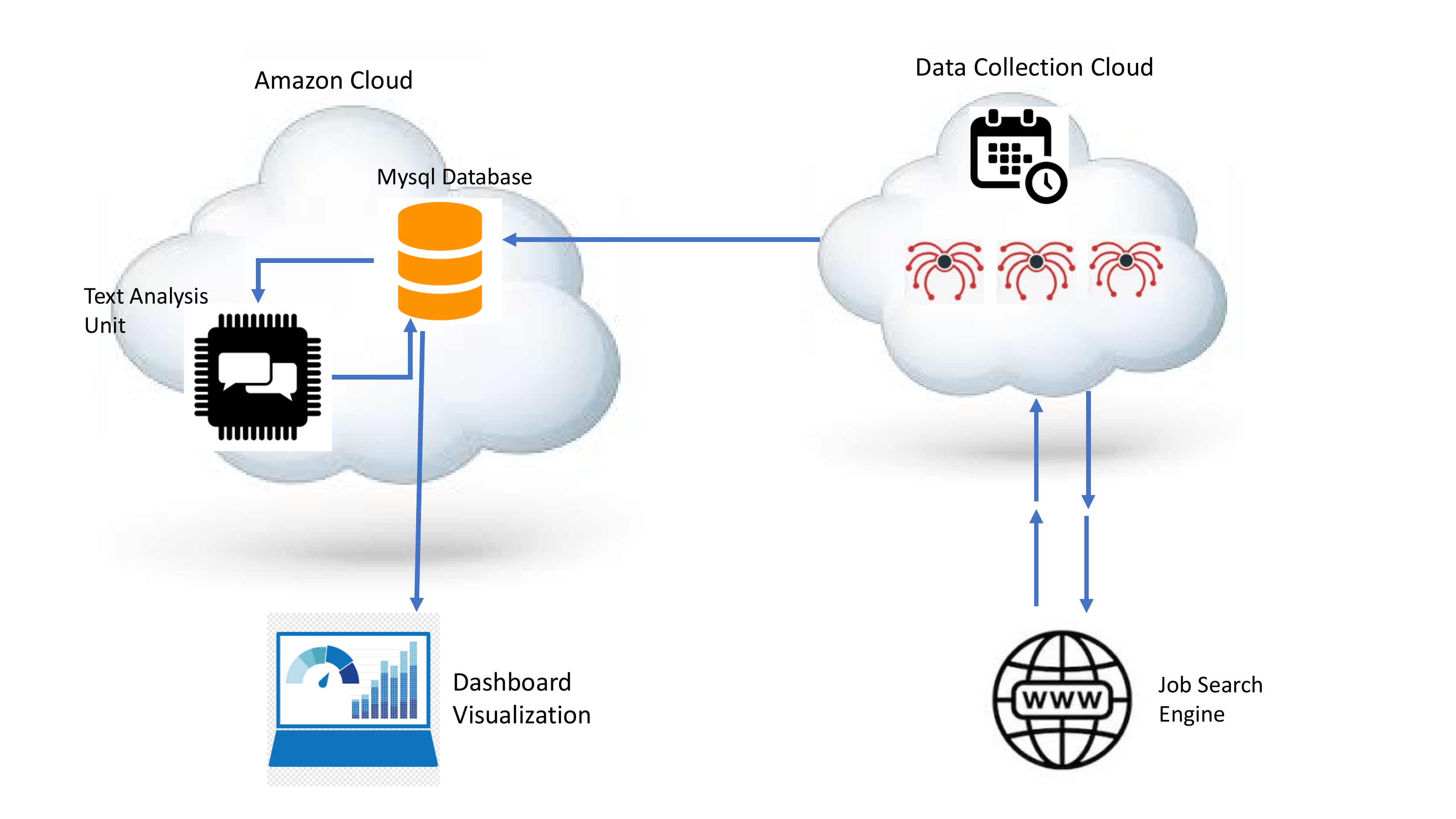}

\caption{Overview of data science skill tracking system}\label{Fig_schema1}
\end{figure*}
Another popular method for this task is using occupational taxonomies that are created by organizations such as ESCO, ISCO or O*NET ~\cite{burrus2013identifying}. As an example, Karakatsanis et al. used O*NET occupation descriptions and job postings while applying Latent Semantic Indexing~(LSI) to both and using cosine similarity between the results~\cite{karakatsanis2017data}. other researchers used techniques such as LDA or word2vec to classify the job postings into ISCO or O*NET classes~\cite{colace2019towards,colombo2018applying}. The problem with these taxonomies is the constant need for updates by experts~\cite{djumalieva2018open}. However, depending on the quality of the generated lexicon, this approaches will yield acceptable results. Other researchers used algorithmic approaches such the work of Mirjana Pejic-Bacha et al. where they use phrase clustering with Jaccard coefficient for distances on industry 4.0 job advertisements~\cite{pejic2020text}. Other researchers used innovative methods such as ontology-based or graph-based models(using hyperlink Wikipedia graph) to extract skills~\cite{sibarani2017ontology,kivimaki2013graph}.
For our implementation on this framework, we build an external lexicon specific to data science skills using information on the web that will cover most of skills and concepts of this field for further analysis. In the next section the details of this approach will be presented.

\section{Method}\label{method}\label{sec:method}
The framework presented in this work consists of multiple sub-systems. Mainly, there are three modules that should interact with each other seamlessly by passing relevant data. The first module is responsible for data collection, the second module processes the collected data, in this case job descriptions from the web and finally, the last module will provide an interface to the insights gathered and provides visualizations for users to examine temporal and spatial distribution of data science related jobs and skills. Figure~\ref{Fig_schema1} shows an overview of the proposed system.
Each of these modules will be described in more detail in the following. As it can be seen in Figure~\ref{Fig_schema1}
, this system heavily relies on cloud computing structures that provide stable facilities for continuous tasks with scheduling to collect, analyze, and present its findings dynamically. 

\subsection{Data Collection Module}

The first module of this framework handles the data collection for the system. The purpose of this unit is to continuously collect a consistent sample of job advertisements from the web. To collect the job posts, the module uses Scrapy which is a very powerful python Framework able to handle different aspects of data collection by its object-oriented implementation using spiders. Spiders are web crawlers that follow the links and can parse the result and extract the requested part at the same time.

These spiders return the results of queries for the following terms: ``data scientist'', ``data analyst'', and ``machine learning engineer'' in job search engine. These three queries were selected based on the main tracks that individual want to pursue in this field. Nevertheless, the returned results covers a very broad spectrum of job positions including more specific jobs such as AI, Computer Vision, or NLP specialist to name a few. In addition, the results not only include the job descriptions which are used for skill extraction but also includes metadata from job advertisement posts allowing temporal and spatial mapping of the result of queries for final visualization. Table~\ref{ta:sample} shows samples of the returned fields for queries on data analyst. 
\begin{table}[b]
\centering
\caption{data collection sample}
\label{ta:sample}
\resizebox{1\textwidth}{!}{%
\begin{tabular}{|c|c|c|c|c|c|c|}
\hline
job\_title                  & company                                      & description                                                        & posted\_date & state & city       & term         \\ \hline
Analytics Analyst II        & Horizon Blue Cross                           & job summary:...                                                    & 4/25/2020    & NJ    & Hopewell   & data analyst \\ \hline
\begin{tabular}[c]{@{}l@{}}Sr. Customer \\Data Analyst  \end{tabular} & Bottomline Technologies                      & bottomline is...                                                   & 4/25/2020    & NH    & Portsmouth & data analyst \\ \hline
\begin{tabular}[c]{@{}l@{}}Master Data \\ Project Analyst \end{tabular} & BaronHR Staffing                             & hiring experience...                                               & 4/25/2020    & TX    & Plano      & data analyst \\ \hline
Business Analyst            & AltaSource Group                             & \begin{tabular}[c]{@{}l@{}}consultant-\\business analyst ...    \end{tabular}                               & 4/25/2020    & WA    & Seattle    & data analyst \\ \hline
Business Analyst            & FHLB Office of Finance                       & \begin{tabular}[c]{@{}l@{}}position:\\business analyst….    \end{tabular}                                   & 4/25/2020    & VA    & Reston     & data analyst \\ \hline
\end{tabular}%
}
\end{table}
For a framework like the one presented, it is not feasible to use a local machine for data collection as it requires to work continuously for a long period of time with no interruptions (such as internet disconnection or power outage, etc.). Such a machine and maintaining its connectivity would cause a lot of hassle and scrapy cloud services solve this problem by understanding the data collection framework. It allows to create periodic jobs to continuously collect the data using this platform. This module runs its data collection spiders every week and samples job postings on web periodically. The results of these queries are then delivered into a mysql database on Amazon web services (AWS) for further processing that will be explained in the next section.

\subsection{Skill Extraction Module}\label{subsect:text_module}

The second module of this framework is the skill extraction module that process the data collected by data collection module. Similar to the previous step, this task should also be continuous and thus resides on the cloud. This work uses AWS architecture as a solution for continuous storage of data and text processing for extracting skills. Amazon Web Services is a safe and well-developed solution for cloud computing and storage. By connecting the output of scrapy cloud service to the amazon databases, we collect data and store them as tables in an AWS mysql database. 
\begin{figure}[t]

\centering
\includegraphics[width=0.85\columnwidth]{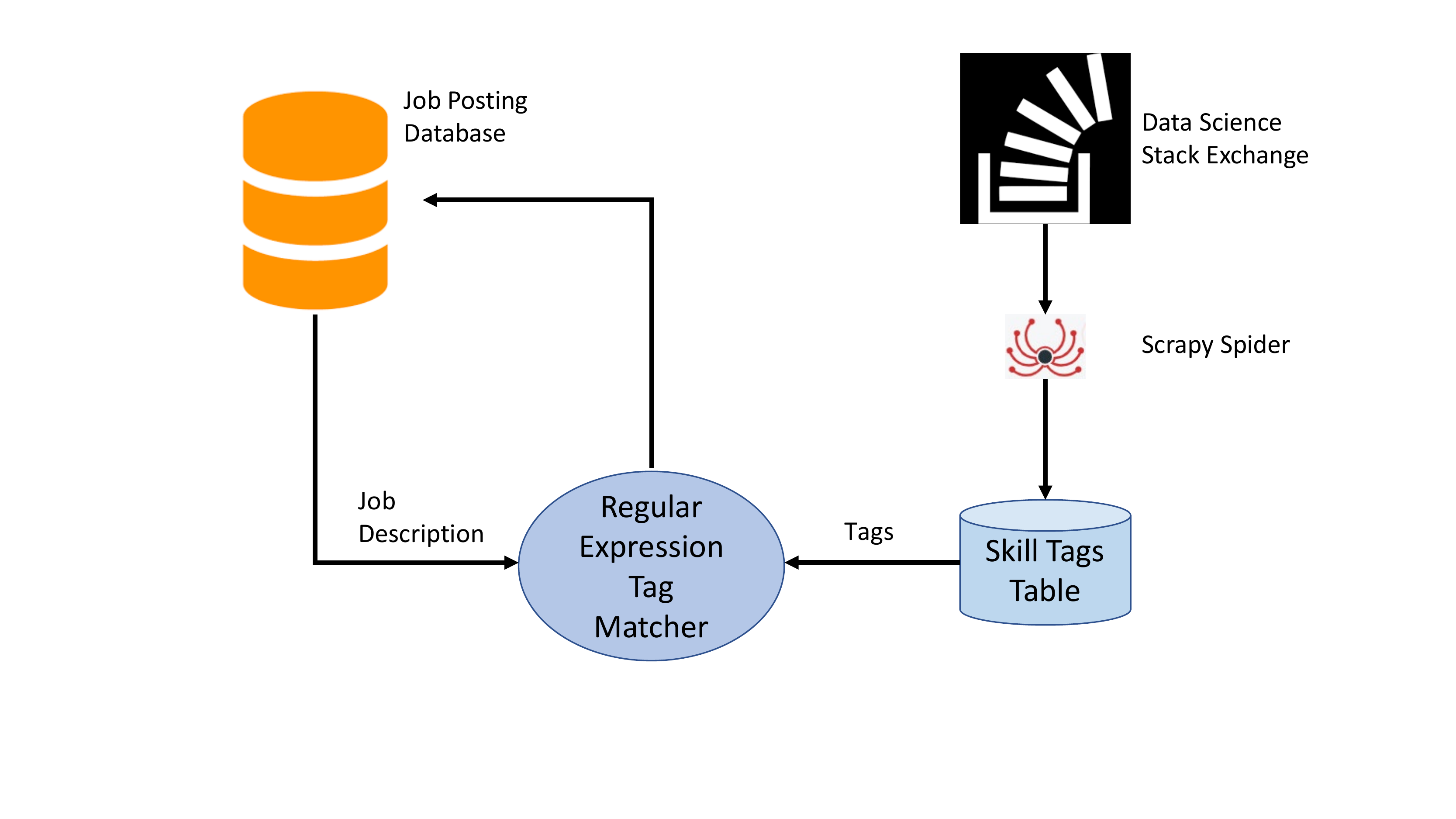}
\vspace{-10pt}
\caption{Overview of skill extraction mechanism }\label{fig:text-unit}
\end{figure}
The Text Analysis Unit is using AWS Lambda service  in combination with cloud watch service. Clould watch allows specifying a periodical task by defining a cron expression. As a result, the python script will be executed periodically based on the schedule given in the form of cron expression. The module perform skill extraction on the collected data and add the result to the database for later usage by the visualization module.
\begin{figure}[!b]
\centering

\includegraphics[width=0.55\textwidth]{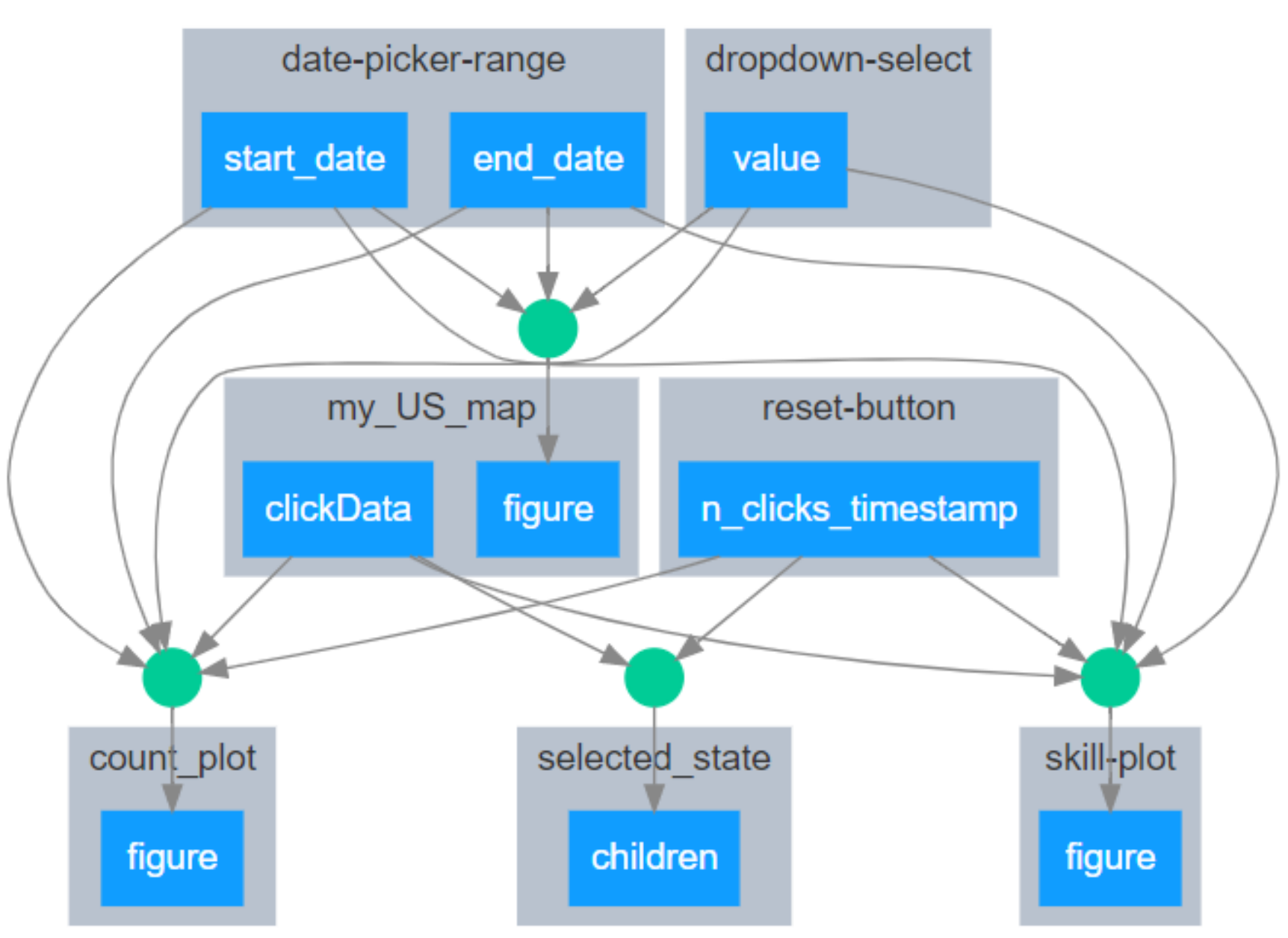}
\caption{Interaction graph of visualization components }\label{fig:interact}
\end{figure}
Given that job posts often include required skills, skill extraction of them would give us valuable insights as to what skills the job market mainly looks for. To extract the skills, an external lexicon has been developed by using \url{datascience.stackexchange.com}. The website is a platform for asking questions related to topics within the data science community along with proper tags. Such tags would cover a large set of concepts and skills needed for data science tasks. Figure~\ref{fig:text-unit} shows the text analysis unit architecture. First, 
a spider crawled the website and all labels of the website have been collected. The lookup table of skills generated then will be fed into a regular expression pattern-matcher to find matches in the job descriptions. The result of this match will be added to each description in the database table. Doing so will identify a majority of the main skills mentioned in the job market given the large number of words in this lexicon. Next, we will discuss the visualization module that is using the data processed by this module.

\subsection{Web-based Visualization}\label{subsec:Web-baseVisualization}

A Framework such as the one presented here, requires a way to present insights by visualization. However due to complexity of the interactions between fields a simple static presentation of results would not provide significant abilities to investigate temporal and spatial components of the job market in the United State. Therefore, a dynamic visualization that provides a mean to interact with different components (e.g. time range and states) would be preferred. Common solutions to provide a dynamic visualization includes R Shiny, and plotly dash for python. Plotly dash is a powerful framework which is build on top of flask framework and allows interaction with the created dashboard for users. It is also compatible with python and thus was used for the app with dynamic visualization in our work. The app has 6 main components:
\begin{itemize}
    \item Job title selection: allowing user to select any of the three main branches of data science or all of them as the input
    \item Data range: allows selecting a period by user.
    \item US map: present the result of the input over the US map and allows selecting a specific state for further investigation.
    \item Skills visualization: presents top skills given the other fields.
    \item Count visualization: number of jobs samples during each week given the period and the state
    \item Reset state: resent state selection and returns the states to the United State as a whole.
\end{itemize}
The effect and interactions between these components are complex. Figure~\ref{fig:interact} shows these interactions in our dash model. As it can be seen the spatial and temporal aspect of skills could be investigated by using these components. Other components would be added as the work continues.

\begin{table}[!b]
\caption{Top 20 results based on no. job by states and company}
\begin{subtable}[h]{0.5\textwidth}
\centering
\caption{Top 20 companies based on no. job posting}\label{ta:companie}
\label{ta:companies}
\begin{tabular}{cl}
\hline
Company                        & Total Ads \\ \hline
Amazon.com Services LLC        & ~~~7,251        \\ \hline
JPMorgan Chase Bank, N.A.      & ~~~3,873        \\ \hline
Deloitte                       & ~~~3,461        \\ \hline
Wells Fargo                    & ~~~2,030        \\ \hline
Amazon Web Services, Inc.      & ~~~1,850        \\ \hline
Microsoft                      & ~~~1,661        \\ \hline
Accenture                      & ~~~1,474        \\ \hline
Pearson                        & ~~~1,388        \\ \hline
Booz Allen Hamilton            & ~~~1,266        \\ \hline
Apple                          & ~~~1,243        \\ \hline
Capital One - US               & ~~~1,166        \\ \hline
US Department of the Air Force & ~~~1,103        \\ \hline
CACI                           & ~~~1,084        \\ \hline
Amazon Dev Center U.S., Inc.   & ~~~1,082        \\ \hline
Facebook                       & ~~~1,069        \\ \hline
Thermo Fisher Scientific       & ~~~1,037        \\ \hline
US Department of the Navy      & ~~~999         \\ \hline
US Department of the Army      & ~~~969         \\ \hline
UnitedHealth Group             & ~~~967         \\ \hline
Capital One                    & ~~~950         \\ \hline
\end{tabular}~~~~\end{subtable}
 \hfill
\begin{subtable}[h]{0.4\textwidth}
\centering
\caption{Top 20 states based on no. job posting}
\label{ta:states}
\begin{tabular}{cl}
\hline
State                       &~~~~ No. Job ads \\ \hline
California                  &~~~~ 35.4K         \\ \hline
Texas                       &~~~~ 17.3K       \\ \hline
Virginia                    &~~~~ 16K         \\ \hline
New York                    &~~~~ 15.7K          \\ \hline
Washington                  &~~~~ 15K        \\ \hline
Illinois                    &~~~~ 11K          \\ \hline
Massachusetts               &~~~~ 10.5K        \\ \hline
Florida                     &~~~~ 8.8K          \\ \hline
pennsylvania                &~~~~ 8.5K          \\ \hline
Maryland                    &~~~~ 8.2K        \\ \hline
North Carolina              &~~~~ 8.1K         \\ \hline
Georgia                     &~~~~ 7.4K       \\ \hline
New Jersey                   &~~~~ 7.3K         \\ \hline
Ohio                    &~~~~ 5.9K          \\ \hline
Colorado                  &~~~~ 5.7K        \\ \hline
Minnesota                    &~~~~ 4.8K          \\ \hline
Michigan               &~~~~ 4.5K        \\ \hline
Arizona                     &~~~~ 4.5K          \\ \hline
Missouri                &~~~~ 3.7K          \\ \hline
Tennessee                    &~~~~ 3.2K        \\ \hline
\end{tabular}~~~~
\end{subtable}
\end{table}


\section{Results}\label{sec:results}
In this section we present our results based on a 12 month period of April 2020 to April 2021. The framework collected approximately 244K job postings. This includes 129K, 71K, 43K job posting for data analyst, machine learning engineering, and data scientist respectively. This shows the job postings under the query term `` data scientist'' are significantly less than the other two categories. Similarly, to build or lexicon, 612 labels were collected covering a broad spectrum of skills and concepts related to data science field using the method described in section~\ref{sec:method}.

\subsection{Aggregated Results}
First, the spatial aspect of job ads in the united states has been investigated. Looking at the aggregated results, the top states with the most job postings in this period can be extracted. Table ~\ref{ta:states} shows the top 20 states according to the number of job posted in this period. The result given in this table is aggregated for all job categories but the framework interface allows to investigate the results by job title and period of time.

\begin{table}[t]
\caption{Top skills for the three main tracks}
\label{ta:skills}
{\fontsize{8}{10pt}
\selectfont
\begin{subtable}[h]{0.29\textwidth}
\caption{Data Scientist}
\begin{tabular}{cl}
\hline
Python                  &$25.2K $          \\ \hline
Machine learning        &$22.8K $         \\ \hline
Statistics              &$21.2K $          \\ \hline
SQL                     &$18.2K $         \\ \hline
Progeramming            &$17.0K   $          \\ \hline
R                       &$16.6K $         \\ \hline
Mathematics             &$1.05K   $          \\ \hline
Algorithms              &$13.1K $        \\ \hline
Data analysis           &$10.9K $          \\ \hline
Visualization           &$9.9K  $         \\ \hline
Cloud                   &$9.7K  $          \\ \hline
AI                      &$8.1K  $         \\ \hline
SAS                     &$8.1K  $          \\ \hline
Databases               &$7.9K  $         \\ \hline
Tableau                 &$7.7K  $          \\ \hline
Spark                   &$7.3K  $          \\ \hline
AWS                     &$7.3K  $          \\ \hline
Deep Learning           &$7.3K  $          \\ \hline
Java                    &$7.0K    $          \\ \hline
Hadoop                  &$5.8K  $         \\ \hline
\end{tabular}
\end{subtable}\hfill
\begin{subtable}[h]{0.39\textwidth}
\caption{Machine Learning Engineer}
\begin{tabular}{cl}
\hline
 Machine learning              &$ 60.7K  $      \\ \hline
 Python                        &$ 36.0K    $      \\ \hline
 Cloud                         &$ 30.0K    $      \\ \hline
 Progeramming                  &$ 27.4K  $      \\ \hline
 Java                          &$ 24.5K  $      \\ \hline
 AWS                           &$ 22.0K    $      \\ \hline
 Software Development          &$ 21.1K  $      \\ \hline
 Algorithms                    &$ 19.5K  $      \\ \hline
 SQL                           &$ 17.4K  $      \\ \hline
 AI                            &$ 15.6K  $      \\ \hline
 C++                           &$ 13.5K  $      \\ \hline
 Spark                         &$ 11.8K  $      \\ \hline
 Databases                     &$ 11.3K  $      \\ \hline
 Statistics                    &$ 9.8K   $      \\ \hline
 Mathematics                   &$ 9.5K   $      \\ \hline
 Linux                         &$ 9.5K   $      \\ \hline
 Optimization                  &$ 9.5K   $      \\ \hline
 Javascript                    &$ 9.0K     $      \\ \hline
 Deep Learning                 &$ 8.8K   $      \\ \hline
 Hadoop                        &$ 8.5K   $      \\ \hline
\end{tabular}
\end{subtable}
\begin{subtable}[h]{0.29\textwidth}
\caption{Data Analyst}
\begin{tabular}{cl}
\hline
 Excel                 &$ 49.7K $     \\ \hline
 SQL                   &$ 43.3K $     \\ \hline
 Data analysis         &$ 28.2K $     \\ \hline
 Tableau               &$ 22.0K   $     \\ \hline
 Databases             &$ 21.0K   $     \\ \hline
 Statistics            &$ 20.0K   $     \\ \hline
 Dashboards            &$ 19.0K   $     \\ \hline
 Visualization         &$ 17.0K   $     \\ \hline
 Programming           &$ 16.9K $     \\ \hline
 Python                &$ 16.2K $     \\ \hline
 R                     &$ 12.3K $     \\ \hline
 Mathematics           &$ 11.6K $     \\ \hline
 Cloud                 &$ 9.8K  $     \\ \hline
 SAS                   &$ 9.5K  $     \\ \hline
 Forcasting            &$ 8.4K  $     \\ \hline
 ETL                   &$ 7.7K  $     \\ \hline
 Data mining           &$ 7.1K  $     \\ \hline
 SAP                   &$ 6.0K    $     \\ \hline
 Machine learning      &$ 5.0K    $     \\ \hline
 Classification        &$ 5.2K  $     \\ \hline
\end{tabular}%
\end{subtable}
}

\end{table}

Furthermore, by aggregating the skills in job postings and looking at the most repeated skills, we could identify the top mentioned skills for each category and the differences between them. Table~\ref{ta:skills} shows the top 20 skills in each job posting category. The result in this table clearly shows the difference between these main tracks of data science. While machine learning engineer job postings mostly shows a domination of skills such as programming, machine learning, cloud, and big data technologies, a data analyst mostly needs skills for data retrieving (e.g. SQL, Excel,database), and Visualization (e.g Tableau, Power BI). A candidate for data scientist jobs, on the other hand would need skills from both of the previous categories. Other important aspect that this result show points to the importance of deep learning, cloud, and big data knowledge for a data scientist. Another interesting observation here is that a common skill for all of those categories is python which appears more than R suggesting that it is a more popular language for data related jobs in industry by comparison.

Lastly, We investigated the top companies that posted jobs related to data science in this period. Table~\ref{ta:companie} shows the top 20 companies during this period. Despite the expectation that tech sector dominate the field, it appears that other sectors contribute to this market significantly. Namely, consultant and advisory section (e.g. Booz Allen, Deloitte)and financial institutes(e.g JPMorgan, Wells Fargo) and government organization contribute significantly to this market along with the tech sector.
\subsection{Temporal Insights}
\begin{figure}[t]
     \centering
     \begin{subfigure}[b]{0.49\textwidth}
         \centering
         \includegraphics[width=\textwidth]{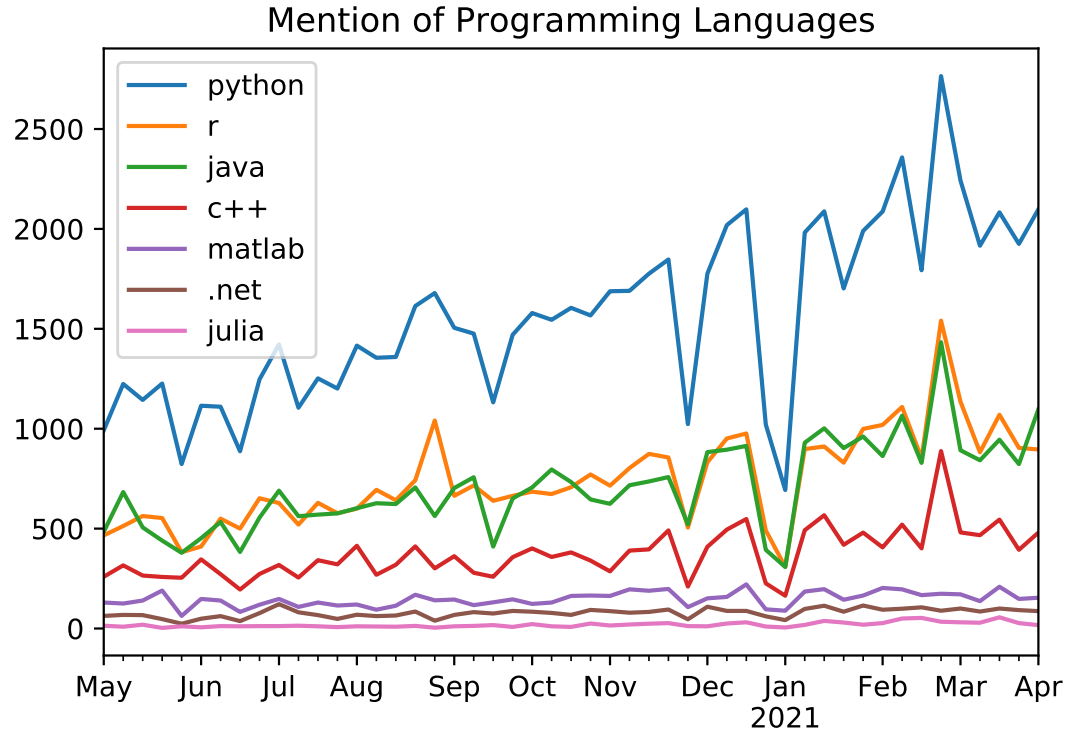}
         \caption{weekly comparison of programming languages}
         \label{fig:prog}
     \end{subfigure}
     \hfill
     \begin{subfigure}[b]{0.49\textwidth}
         \centering
         \includegraphics[width=\textwidth]{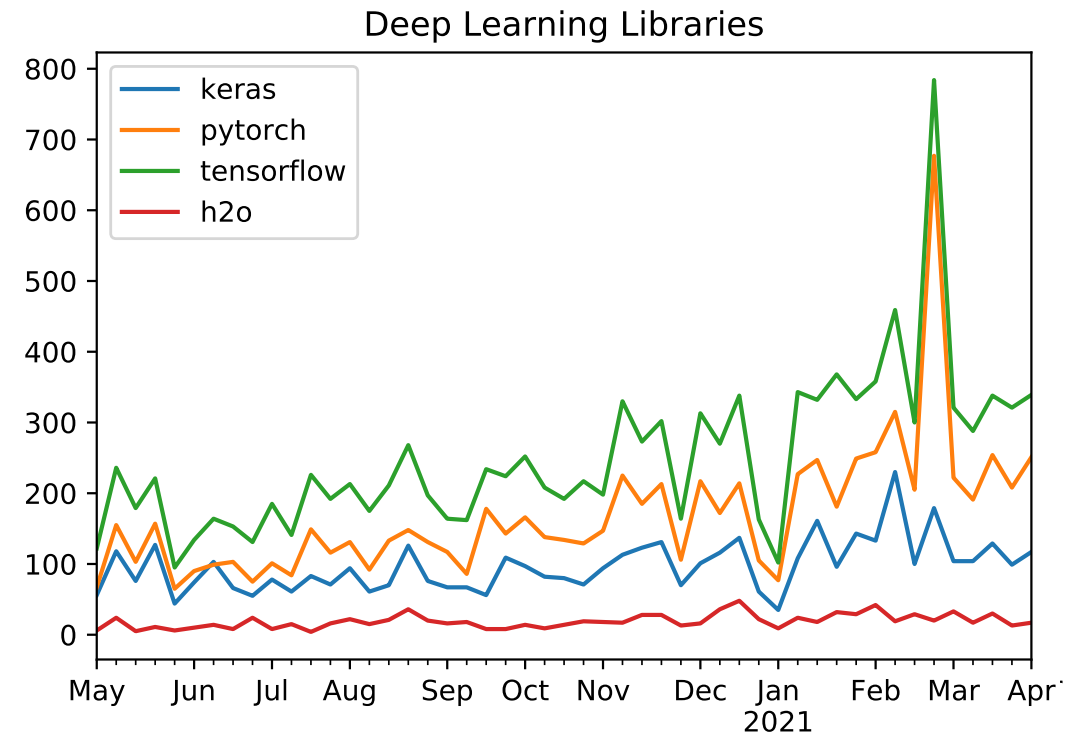}
         \caption{weekly comparison of deep learning frameworks}
         \label{fig:dp}
     \end{subfigure}
        \caption{Temporal insights of data science skills}
        \label{fig:insights}
\end{figure}
The framework allows us to investigate the interested skills along the time. Certain comparisons is being presented as a result of this system to provide extra insights that this system can provide. One important insight is the contribution and importance of programming languages required for data scientist. Figure~\ref{fig:prog} shows this comparison between top programming languages mentioned during this period.As it can be seen, the top language is python following by R and Java. This indicated not only python is the most dominant language but the trend and the gap shows it appears to be in demand in future in comparison to others. Similarly, a comparison between deep learning framework is presented in Figure~\ref{fig:dp}. The result shows that tensorflow followed by pytorch are the most mentioned deep learning frameworks. Also H2O~(an R based package) is mentioned much less which indicates the overwhelming attention to python for deep learning. As it can be seen, the framework allows temporal insights to this job market and based on their general importance, such insights can be added to the interface designed for users.
\begin{figure*}[t]
\centering
\includegraphics[width=\textwidth]{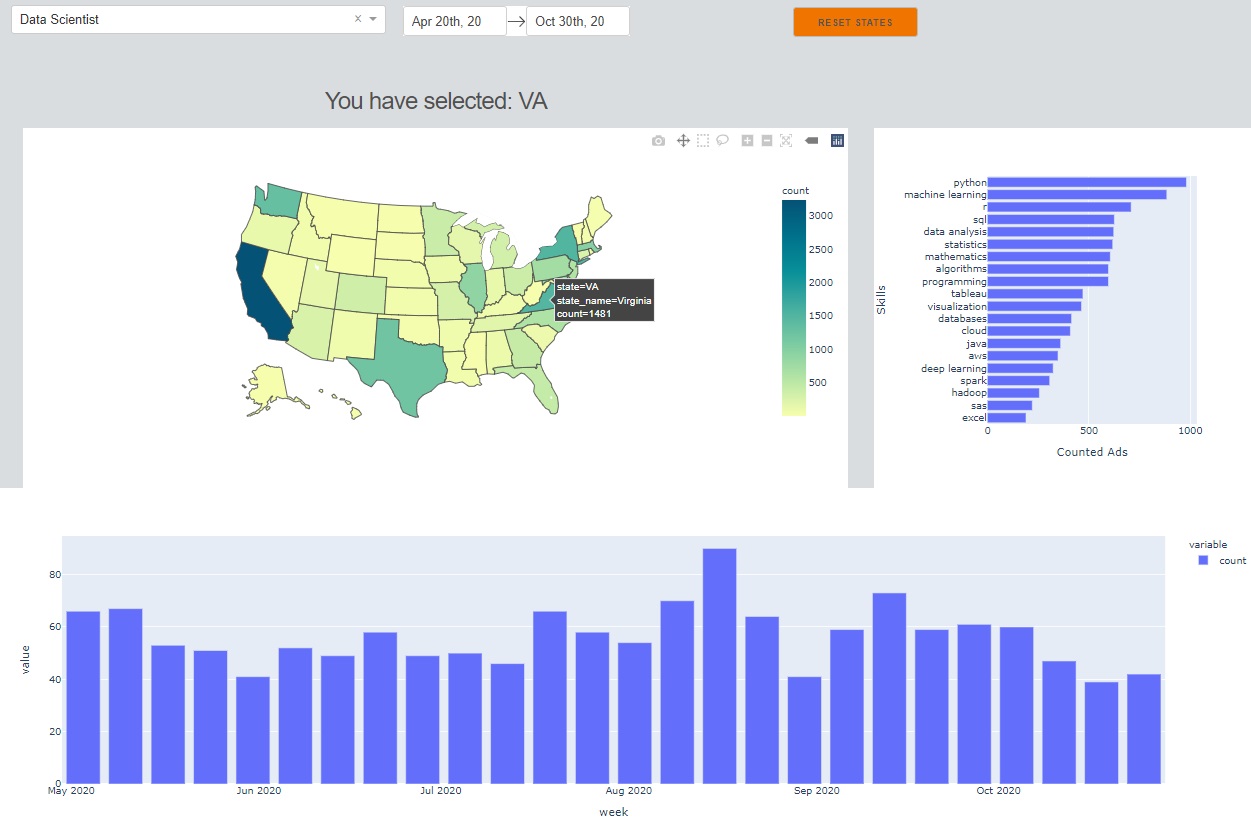}
\caption{Web-based interface of job market monitoring app }\label{fig: result}
\end{figure*}
Finally, the framework provides an interface for individuals to interact with. The interface deployed on heroku as host for this application using 1 dyno at the moment with the latest update in April (available at~\url{dsi-usa2.herokuapp.com}). The interface connects to AWS database and retrieves relevant pieces of information for the visualization. Figure~\ref{fig: result} shows the interface as a standalone application. Users can select any query field in a given period of time and the skillset (counts) and weekly number of advertise will be updated accordingly. Also, selecting any state will  update these charts accordingly. To return back to all of the United States, ``reset states'' button should be used. The weekly count chart allows us to track the number of job postings during a period.

\section{Conclusion and Future Work}\label{sec:Conclusion}
This paper introduces a framework capable of collecting and analysing of spatial and temporal aspects of data science's job market using available tools within data science toolbox. The framework provides a free front-end interface developed using plotly dash for this market. The result provides a quantitative guide for individuals and organizations to recognize the most important skills and concepts in this domain based on industry perception.
Possible extension to the work here includes sector analysis of job posting and improving on skill extraction technique. Other temporal insights will be added to the interface based on the importance of the information they can add for individuals using this interface.

\bibliographystyle{bibtex/spmpsci} 
\bibliography{main.bib}

\begin{thebibliography}{10}
\providecommand{\url}[1]{{#1}}
\providecommand{\urlprefix}{URL }
\expandafter\ifx\csname urlstyle\endcsname\relax
  \providecommand{\doi}[1]{DOI~\discretionary{}{}{}#1}\else
  \providecommand{\doi}{DOI~\discretionary{}{}{}\begingroup
  \urlstyle{rm}\Url}\fi

\bibitem{linkedin2017}
Linkedin’s 2017 u.s. emerging jobs report.
\newblock
  \url{https://economicgraph.linkedin.com/research/LinkedIns-2017-US-Emerging-Jobs-Report}.
\newblock Accessed: 2020-03-29

\bibitem{belloum2019bridging}
Belloum, A.S., Koulouzis, S., Wiktorski, T., Manieri, A.: Bridging the demand
  and the offer in data science.
\newblock Concurrency and Computation: Practice and Experience \textbf{31}(17),
  e5200 (2019)

\bibitem{burrus2013identifying}
Burrus, J., Jackson, T., Xi, N., Steinberg, J.: Identifying the most important
  21st century workforce competencies: An analysis of the occupational
  information network (o* net).
\newblock ETS Research Report Series \textbf{2013}(2), i--55 (2013)

\bibitem{carmichael2018data}
Carmichael, I., Marron, J.: Data science vs. statistics: two cultures?
\newblock Japanese Journal of Statistics and Data Science \textbf{1}(1),
  117--138 (2018)

\bibitem{colace2019towards}
Colace, F., De~Santo, M., Lombardi, M., Mercorio, F., Mezzanzanica, M.,
  Pascale, F.: Towards labour market intelligence through topic modelling.
\newblock In: Proceedings of the 52nd Hawaii International Conference on System
  Sciences (2019)

\bibitem{colombo2018applying}
Colombo, E., Mercorio, F., Mezzanzanica, M.: Applying machine learning tools on
  web vacancies for labour market and skill analysis.
\newblock Terminator or the Jetsons? The Economics and Policy Implications of
  Artificial Intelligence  (2018)

\bibitem{daneva2017job}
Daneva, M., Wang, C., Hoener, P.: What the job market wants from requirements
  engineers? an empirical analysis of online job ads from the netherlands.
\newblock In: 2017 ACM/IEEE International Symposium on Empirical Software
  Engineering and Measurement (ESEM), pp. 448--453. IEEE (2017)

\bibitem{darabi2018detecting}
Darabi, H., Karim, F., Harford, S., Douzali, E., Nelson, P.: Detecting current
  job market skills and requirements through text min-ing.
\newblock In: 2018 ASEE Annual Conference and Exposition, Conference
  Proceedings (2018)

\bibitem{davenport2012data}
Davenport, T.H., Patil, D.: Data scientist.
\newblock Harvard business review \textbf{90}(5), 70--76 (2012)

\bibitem{dhar2013data}
Dhar, V.: Data science and prediction.
\newblock Communications of the ACM \textbf{56}(12), 64--73 (2013)

\bibitem{djumalieva2018open}
Djumalieva, J., Sleeman, C.: An open and data-driven taxonomy of skills
  extracted from online job adverts.
\newblock Developing Skills in a Changing World of Work: Concepts, Measurement
  and Data Applied in Regional and Local Labour Market Monitoring Across Europe
  \textbf{425} (2018)

\bibitem{gardiner2018skill}
Gardiner, A., Aasheim, C., Rutner, P., Williams, S.: Skill requirements in big
  data: A content analysis of job advertisements.
\newblock Journal of Computer Information Systems \textbf{58}(4), 374--384
  (2018)

\bibitem{hartnett2014nasig}
Hartnett, E.: Nasig's core competencies for electronic resources librarians
  revisited: An analysis of job advertisement trends, 2000--2012.
\newblock The journal of academic librarianship \textbf{40}(3-4), 247--258
  (2014)

\bibitem{karakatsanis2017data}
Karakatsanis, I., AlKhader, W., MacCrory, F., Alibasic, A., Omar, M.A., Aung,
  Z., Woon, W.L.: Data mining approach to monitoring the requirements of the
  job market: A case study.
\newblock Information Systems \textbf{65}, 1--6 (2017)

\bibitem{kivimaki2013graph}
Kivim{\"a}ki, I., Panchenko, A., Dessy, A., Verdegem, D., Francq, P., Bersini,
  H., Saerens, M.: A graph-based approach to skill extraction from text.
\newblock In: Proceedings of TextGraphs-8 graph-based methods for natural
  language processing, pp. 79--87 (2013)

\bibitem{maer2019skill}
Maer-Matei, M.M., Mocanu, C., Zamfir, A.M., Georgescu, T.M.: Skill needs for
  early career researchers—a text mining approach.
\newblock Sustainability \textbf{11}(10), 2789 (2019)

\bibitem{manieri2015teaching}
Manieri, A., Nucci, F.S., Femminella, M., Reali, G.: Teaching domain-driven
  data science: public-private co-creation of market-driven certificate.
\newblock In: 2015 IEEE 7th International Conference on Cloud Computing
  Technology and Science (CloudCom), pp. 569--574. IEEE (2015)

\bibitem{miller2017quant}
Miller, S., Hughes, D.: The quant crunch: How the demand for data science
  skills is disrupting the job market.
\newblock Burning Glass Technologies  (2017)

\bibitem{pejic2020text}
Pejic-Bach, M., Bertoncel, T., Me{\v{s}}ko, M., Krsti{\'c}, {\v{Z}}.: Text
  mining of industry 4.0 job advertisements.
\newblock International Journal of Information Management \textbf{50}, 416--431
  (2020)

\bibitem{provost2013data}
Provost, F., Fawcett, T.: Data science and its relationship to big data and
  data-driven decision making.
\newblock Big data \textbf{1}(1), 51--59 (2013)

\bibitem{sibarani2017ontology}
Sibarani, E.M., Scerri, S., Morales, C., Auer, S., Collarana, D.:
  Ontology-guided job market demand analysis: a cross-sectional study for the
  data science field.
\newblock In: Proceedings of the 13th International Conference on Semantic
  Systems, pp. 25--32 (2017)

\bibitem{wowczko2015skills}
Wowczko, I.A.: Skills and vacancy analysis with data mining techniques.
\newblock In: Informatics, vol.~2, pp. 31--49. Multidisciplinary Digital
  Publishing Institute (2015)

\bibitem{yang2016current}
Yang, Q., Zhang, X., Du, X., Bielefield, A., Liu, Y.Q.: Current market demand
  for core competencies of librarianship—a text mining study of american
  library association’s advertisements from 2009 through 2014.
\newblock Applied Sciences \textbf{6}(2), 48 (2016)

\end{thebibliography}
%
%

\end{document}